# High-brightness switchable multi-wavelength remote laser in air

Jinping Yao[1], Bin Zeng[1,3], Huailiang Xu[2,†], Guihua Li[1,3], Wei Chu[1,3], Jielei Ni[1,3], Haisu Zhang[1,3], See Leang Chin[4], Ya Cheng[1,‡], and Zhizhan Xu[1,★]

[1] State Key Laboratory of High Field Laser Physics, Shanghai Institute of Optics and Fine Mechanics, Chinese Academy of Sciences, Shanghai 201800, China

[2] State Key Laboratory on Integrated Optoelectronics, College of Electronic Science and Engineering, Jilin University, Changchun 130012, China

[3] Graduate School of Chinese Academy of Sciences, Beijing 100080, China

[4] Center for Optics, Photonics and Laser (COPL) & Department of Physics, Engineering Physics and Optics, Université Laval, Quebec City, Qc G1V 0A6, Canada

[†] Email: huailiang@jlu.edu.cn
[‡] Email: ycheng-45277@hotmail.com
[★] Email: zzxu@mail.shcnc.ac.cn

**Abstract**

Remote laser in air based on amplified spontaneous emission (ASE) has produced rather well-collimated coherent beams in both backward and forward propagation directions [1-3], opening up possibilities for new remote sensing approaches [4,5]. The remote ASE-based lasers were shown to enable operation either at ~391 and 337 nm using molecular nitrogen or at ~845 nm using molecular oxygen as gain medium, depending on the employed pump lasers. To date, a multi-wavelength laser in air that allows for dynamically switching the operating wavelength has not yet been achieved, although this type of laser is certainly of high importance for detecting multiple hazard gases. In this Letter, we demonstrate, for the first time to our knowledge, a harmonic-seeded switchable multi-wavelength laser in air driven by intense mid-infrared femtosecond laser pulses. Furthermore, population inversion in the multi-wavelength remote laser occurs at an ultrafast time-scale (i.e., less than ~200 fs) owing to direct formation of excited molecular nitrogen ions by strong-field ionization of inner-valence electrons, which is fundamentally different from the previously reported pumping mechanisms based either on electron recombination of ionized molecular nitrogen or on resonant two-photon excitation of atomic oxygen fragments resulting from resonant two-photon dissociation of molecular oxygen. The bright multi-wavelength laser in air opens the perspective for remote detection of multiple pollutants based on nonlinear spectroscopy [6].

**PACS numbers**: 42.65.Re

Since the first ruby laser was demonstrated in 1960, there have been enormous requirements for a variety of coherent light sources in a broad spectrum of fields covering science and engineering. In environmental science, there has been a large amount of research effort aiming at measuring atmospheric trace species over a long distance. The ability to control the generation of coherent light source with different frequencies at a designed location would provide a new strategy to meet the pressing needs of various environmental issues from monitoring global warming and stratospheric ozone depletion to early detection of nuclear reactor radiation leak and biological treat agents in air.

Recent advances in femtosecond laser filamentation offer the possibility to generate the switchable multi-wavelength remote laser. When femtosecond laser pulses propagate in air, filamentation can easily occur as a result of the dynamic balance between Kerr self-focusing and plasma defocusing [7-9]. The dynamic balance not only gives rise to an extended spatial confinement of the femtosecond laser beam, but also a high peak intensity stabilized inside the filament core due to the well-known intensity clamping effect [10-12]. Both these effects can be beneficial for enhancing the efficiency of nonlinear optical processes, such as third harmonic generation and four-wave mixing [13-15]. In particular, amplified spontaneous emissions (ASE) of nitrogen molecules at 337 and 391 nm and that of oxygen atoms at 845 nm have been observed in the femtosecond-laser-induced plasma filament in air. In these circumstances, population inversion was ascribed to the recombination of free electrons with molecular nitrogen ions [1,2] and resonant two-photon excitation of atomic oxygen fragments [3]. Although the nitrogen-based remote laser can potentially operate at multiple wavelengths as conventional nitrogen lasers do, switching between different wavelengths by use of optics is difficult owing to the fact that the remote lasing action always occurs far from the pump laser system. We show that this problem can be circumvented by seeding the nitrogen laser in air with self-generated harmonics of an ultrafast wavelength-tunable mid-infrared pump laser.

The remote lasing action was demonstrated with wavelength-tunable mid-infrared laser pulses generated from an optical parametric amplifier (OPA, HE-TOPAS, Light Conversion, Inc.), which was pumped by a commercial Ti:Sapphire laser system (Legend Elite-Duo, Coherent, Inc.). These intense mid-infrared laser pulses have enabled investigation of strong field atomic physics such as above threshold ionization and high-order harmonic generation [16,17]. The central wavelength of the mid-infrared pulses from the OPA can be continuously tuned in the spectral range from ~1200 nm to 2500 nm. In the spectral range of 1600 nm~2500 nm which corresponds to the idler output of OPA, the maximum pulse energy of the OPA could reach ~700 μJ, and the pulse duration around 1900nm was measured to be ~200 fs. The Ti:Sapphire laser, operated at a repetition rate of 1 kHz, provided ~40 fs (FWHM) laser pulses with a central wavelength at ~800 nm and a single pulse energy of ~6 mJ. Tuning wavelength of OPA was achieved by rotating the phase-matched angle of a β-BaB$_2$O$_4$ (BBO) crystal. The pump laser pulses were focused into air using lenses of different focal lengths ranging from 36 mm to 300 mm to generate the 3$^{rd}$ and 5$^{th}$ harmonics as well as the remote laser in air. A grating spectrometer (Shamrock 303i, Andor) with a 1200 grooves/mm grating was used to record the spectra of both neutral and ionized nitrogen molecules. For measuring the fluorescence spectra in the backward or side directions, a fiber head was used to collect the signal into the grating spectrometer. In this case, a plano-convex lens was used to couple the fluorescence light into the fiber head.

Figure 1(a) shows the remote laser spectrum obtained with 1900 nm pump pulses for a pulse energy of ~500 μJ. Unlike the typical supercontinuum-like spectra of the $3^{rd}$ and $5^{th}$ harmonics produced by filamentation of mid-infrared laser pulses [12], a strong, narrow-bandwidth emission at 391 nm appears on-top of the spectrum of the $5^{th}$ harmonic, whose intensity is nearly 2~3 orders of magnitude higher than that of the fluorescence lines at 357 nm and 337nm from neutral molecular nitrogen. In addition, we examined the polarization of the emission at 391 nm using a Glan-Taylor polarizer which was placed just before the spectrometer. As shown in the inset of Fig. 1(a), the strong emission at 391 nm is nearly perfectly linearly polarized in the direction parallel to that of the pump pulse. This provides strong evidence on the harmonic-seeded lasing action at 391 nm because both spontaneous emission and ASE will show isotropic polarization. For comparison, we also present the fluorescence spectra measured for the backward and side emissions, as shown by the solid and dashed curves in Fig. 1(b). Since the seeding effect cannot take place in these directions as the $5^{th}$ harmonic beam always follows the laser propagation in the forward direction, the spectra recorded in the backward and side directions do not show any lasing peak as the 391 nm in Fig. 1 (a); they appear similar to each other as shown in Fig. 1(b) in which only fluorescence lines of comparable intensities from both ionized and neutral nitrogen molecules are observed.

The seeding effect due to the existence of the 5$^{th}$ harmonic in the gain zone is further confirmed by examining the spectra of the remote laser at 391 nm for different pump wavelengths, as shown in Fig. 2(a). When the pump wavelength is tuned to 1906 nm, the intensity of the 391 nm laser is the strongest; however, when the pump wavelength is tuned toward either longer or shorter wavelength, the intensity of the 391 nm laser decreases significantly as shown in Fig. 2(a). The physics behind the dependence of the lasing on the harmonic wavelength will be discussed later. Furthermore, as shown in Fig. 2(b), harmonic-seeded remote lasers can also be realized at multiple wavelengths of 330, 357, 391, 428 and 471 nm which correspond to the transitions between different vibrational levels of the electronic B$^2\Sigma_u^+$ and X $^2\Sigma_g^+$ states of molecular nitrogen ions (see Fig. 5 below), when the pump wavelengths are respectively tuned to 1682, 1760, 1920, 2050 and 1415 nm. It should be noted that since the pump pulse energy of our mid-infrared OPA source drops significantly for the wavelength longer than 2200 nm, the remote lasing action at 471 nm is thus seeded by the 3$^{rd}$ harmonic of the pump pulses, not by the 5$^{th}$ harmonic. However, in this case, since the intensity of the seed is relatively high due to the higher conversion efficiency of the 3$^{rd}$ harmonic, the contrast of the lasing at 471 nm to the 3$^{rd}$ harmonic seed becomes worse. The contrast could be improved with future development of femtosecond OPA technology by which higher pulse energies at wavelengths longer than 2200 nm are attainable.

As a final evidence on the population inversion, we show that the lasing action at 391 nm can also be achieved with externally injected seed pulses offered by the second harmonic of the 800 nm Ti:sapphire laser together with the 1920 nm pump laser pulses. In particular, the second harmonic is polarized in the direction perpendicular to that of the mid-infrared pump pulses, so that it can be easily separated from the self-generated $5^{th}$ harmonic using a Glan-Taylor polarizer. The second harmonic was combined with the 1920 nm pump pulses by use of a dichroic mirror with high reflectivity at 400 nm and high transmission at 1920 nm, and then the collinearly propagating pump pulses at 1920 nm and seed pulses at 400 nm were both focused into air by a plano-convex lens to produce the plasma channel. The temporal synchronization between the second harmonic and the pump pulses was achieved using a time-delay line. The spatial overlap of the two beams was achieved by placing another lens with long focal length in the path of fundamental pulses which compensates the difference between the foci of the second harmonic beam at ~400 nm and the pump beam at ~1920 nm. Figure 3 shows the spectra of the second harmonic measured before (blue dashed curve) and after (red solid curve) switching on the mid-infrared pump laser. Similar to the remote laser seeded with the self-generated $5^{th}$ harmonic, the 391 nm peak appears on the second harmonic spectrum when the pump laser is switched on and temporal and spatial overlap between the second harmonic and the pump laser have been achieved. In addition, when the second harmonic is blocked while the mid-infrared pump laser is still on, only typical fluorescence spectrum of ionized and neutral nitrogen molecules is recorded (i. e., no lasing observed in this case), as shown by the green dotted curve in Fig. 3. Thus, injection of seed pulses is necessary for achieving lasing action.

The gain curve of the remote laser at the wavelength of 391 nm can be obtained by truncating the plasma channel using a pair of uncoated fused silica plates [13] and measuring the energy of the 391 nm laser at each truncation location. Details of the beam truncation optics can be found elsewhere [18]. Both signals of the 391 nm laser and 5$^{th}$ harmonic were recorded using the spectrometer (Shamrock 303i, Andor), and the energy of 391 nm laser was estimated by integrating the spectral flux over a narrow window from 391 nm to 392 nm. The recorded laser spectrum is peaked at 391.4 nm with a spectral width of ~0.3 nm (full width at half maximum, FWHM). To remove the fluorescence signal, a polarizer was placed just before the spectrometer which was set to only allow the light with the polarization direction parallel to that of the pump laser to pass through. Note that the harmonic-seeded laser at 391 nm has the same polarization as that of the pump laser at fundamental wavelength. The measured energy of the 391 nm laser shows clear exponential dependence on the plasma length in Fig. 4 as evidenced by the gain curve with an estimated gain coefficient of ~5.02/cm. This gain coefficient is more than one order of magnitude higher than that of the ASE as reported in Ref. [2], which can be attributed to the seeding effect.

The fact that population inversion could be achieved within an ultrafast time-scale for initiating subsequently amplification of the 3$^{rd}$ and 5$^{th}$ harmonics indicates that the ionization of inner-valence electron has occurred during the strong-field ionization of nitrogen molecules when intense mid-infrared laser fields interact with air molecules. This is because the ejection (ionization) of an inner-valence electron from the $\sigma_u 2s$ orbital of nitrogen molecules leaves the ion in the excited $B^2\Sigma_u^+$ state, whereas the ionization of an outer-valence $\sigma_g 2p$ electron leads to the molecular nitrogen ions lying on the ground $X^2\Sigma_g^+$ state [19-20].

As shown in Fig. 5(a), these are two distinct channels of ionization: one ionizing the outer-valence electron resulting in the ground state of nitrogen molecular ion, whereas the other ionizing the inner-valence electron resulting in an excited ionic state. Figure 5(b) shows the detailed transitions in the second channel. The lower potential curve in this case is empty initially. Thus, any transition from the upper curve to the lower curve would represent that from an inversion system. This mechanism ensures that the population inversion can occur instantly with the ionization of inner-valence electrons, leading to the gain we observed in this channel. On the other hand, the population of the molecular nitrogen ions in the ground state resulting from the other channel could be reduced by absorption of the harmonic photons. As a matter of fact, the lasing wavelengths shown in Fig. 2(b) are located at the long-wavelength side of the harmonic spectra, indicating that the harmonics play an important role in the population inversion. The ground ionic nitrogen molecules will absorb the harmonics which are resonant with some higher vib-rotational levels of the excited $B^2\Sigma_u^+$ state. This will efficiently reduce the population of the molecular nitrogen ions in the ground state, leading to the population inversion for the transitions indicated in Fig. 5. As shown in Fig. 2(a), the observation that the intensity of the generated laser decreases significantly when the pump laser wavelength is tuned off implies that the population inversion condition can be achieved only in certain wavelength range of the pump laser.

This ultrafast population inversion phenomenon is unambiguously confirmed from the observed seeded lasing of nitrogen molecule ions when seeded by the 3rd or 5th harmonics generated by an ultrafast nonlinear process and phase-locked to the fundamental pulses in air [13]. Interestingly, with this harmonic-seeding mechanism, it will be impossible to achieve the seeded lasing action at 337 nm (i. e., corresponding to $C^3\Pi_u - B^3\Pi_g$ transition of neutral nitrogen molecules) even though there is inversion [1,2]. This is because the $C^3\Pi_u$ state of nitrogen molecules results from the dissociative recombination in a picosecond range through the following processes: $N_2^+ + N_2 \Rightarrow N_4^+$; $N_4^+ + e \Rightarrow N_2 (C^3\Pi_u) + N_2$ in air [21]. This prediction is confirmed by the fact that no lasing at 337 nm could be observed in our experiment.

In conclusion, we have demonstrated a remote laser in air that can dynamically switch between different wavelengths. Unlike the ASE-based remote lasers, the multi-wavelength laser seeded by the self-generated harmonics is capable of offering coherent beams of high brightness in both UV and visible regimes. Further development in the mid-infrared ultrafast laser technology will lead to pump laser pulse energies significantly higher than that used in this experiment, by which remote lasers in air with greatly enhanced peak powers and at designed remote locations can be expected. The switchable multi-wavelength remote laser provides a new toolkit for remote nonlinear spectroscopy applications.

The work is supported by National Basic Research Program of China (Grant 2011CB808102), National Natural Science Foundation of China (Grant Nos. 10974213, 60825406, 11074098) and NCET-09-0429. SLC is supported by the Canada Research Chair.

**Figure captions:**

Fig. 1. Typical spectra of nitrogen excited by 1900 nm laser pulses recorded in (a) forward, (b) backward (black solid line) and side (red dashed line) directions. Inset in (a): polarization property of remote laser at ~391 nm. Dashed curves are vertically shifted in (b) for clarity.

Fig. 2. (a) Measured spectra of remote laser at 391 nm for different pump wavelengths of 1834, 1906, and 1960 nm; and (b) the lasing peaks at 471, 428, 391, 357, and 330 nm achieved with different pump wavelengths of 1415, 2050, 1920, 1760, and 1682 nm, respectively. Curves are shifted vertically in (a) for clarity.

Fig. 3. Spectra of externally injected second harmonic of 800 nm Ti:sapphire laser measured before (blue dashed) and after (red solid) switching on the 1920 nm pump laser, and the spectrum obtained with only 1920 nm pump pulses (green dotted).

Fig. 4. Fitted gain curve of harmonic-seeded remote laser in air at ~391 nm.

Fig. 5. (a) Interaction of nitrogen molecules simultaneously with mid-infrared laser field and its 3$^{rd}$ or 5$^{th}$ harmonics resulting in formation of molecular nitrogen ions at excited state by ionization of inner-valance electrons, giving rise to harmonic-seeded laser. (b) Energy level diagram of ionized and neutral nitrogen molecules in which the transitions between $B^2\Sigma_u^+$ and $X^2\Sigma_g^+$ states are indicated with corresponding lasing wavelengths.

Fig. 1

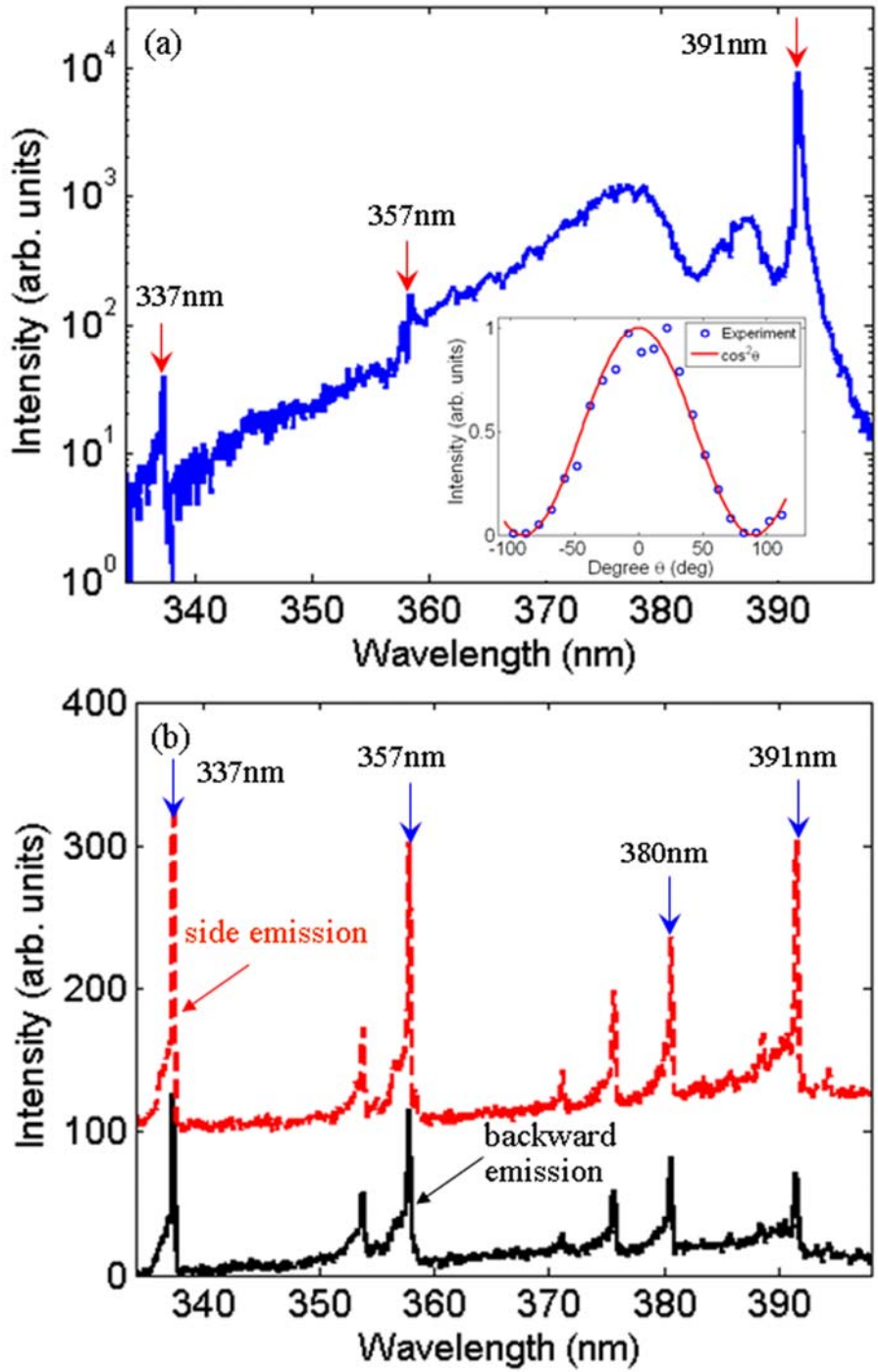

Fig. 2

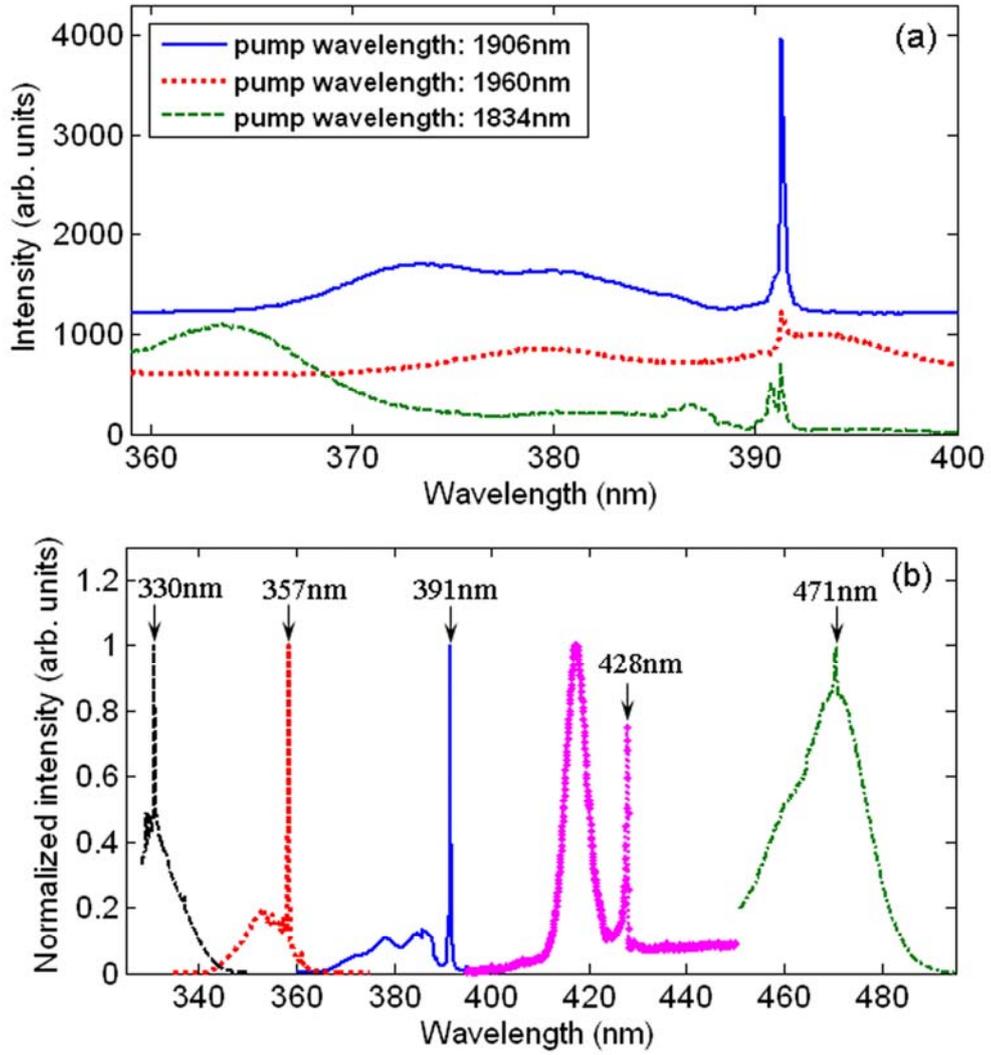

Fig. 3

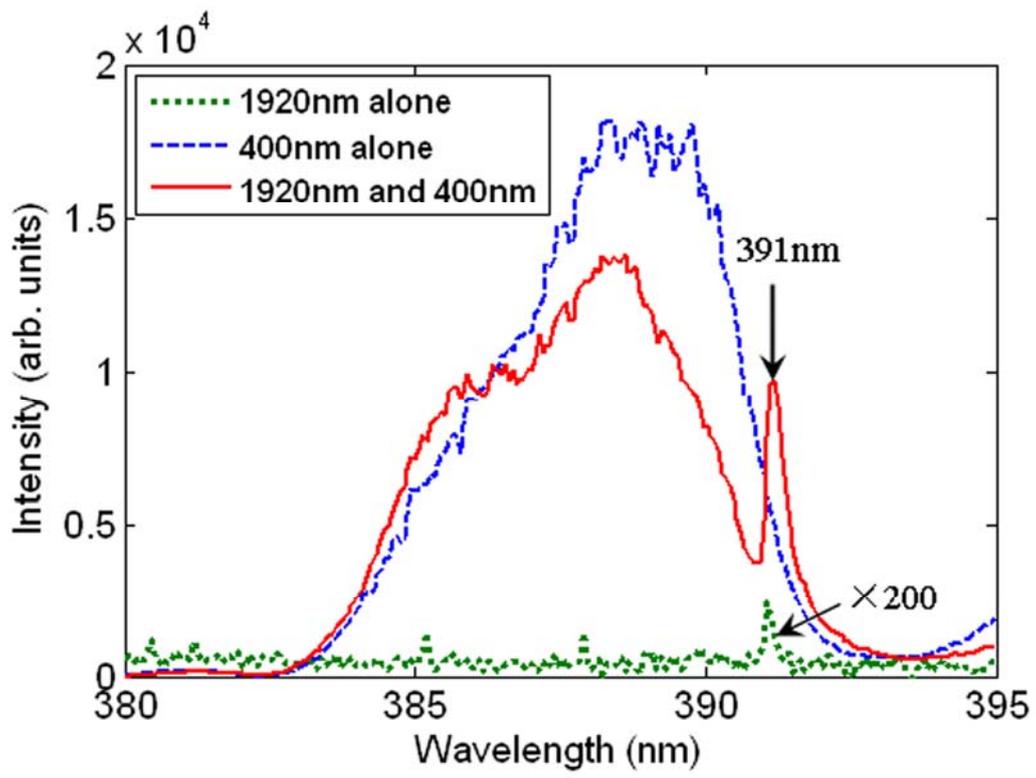

Fig. 4

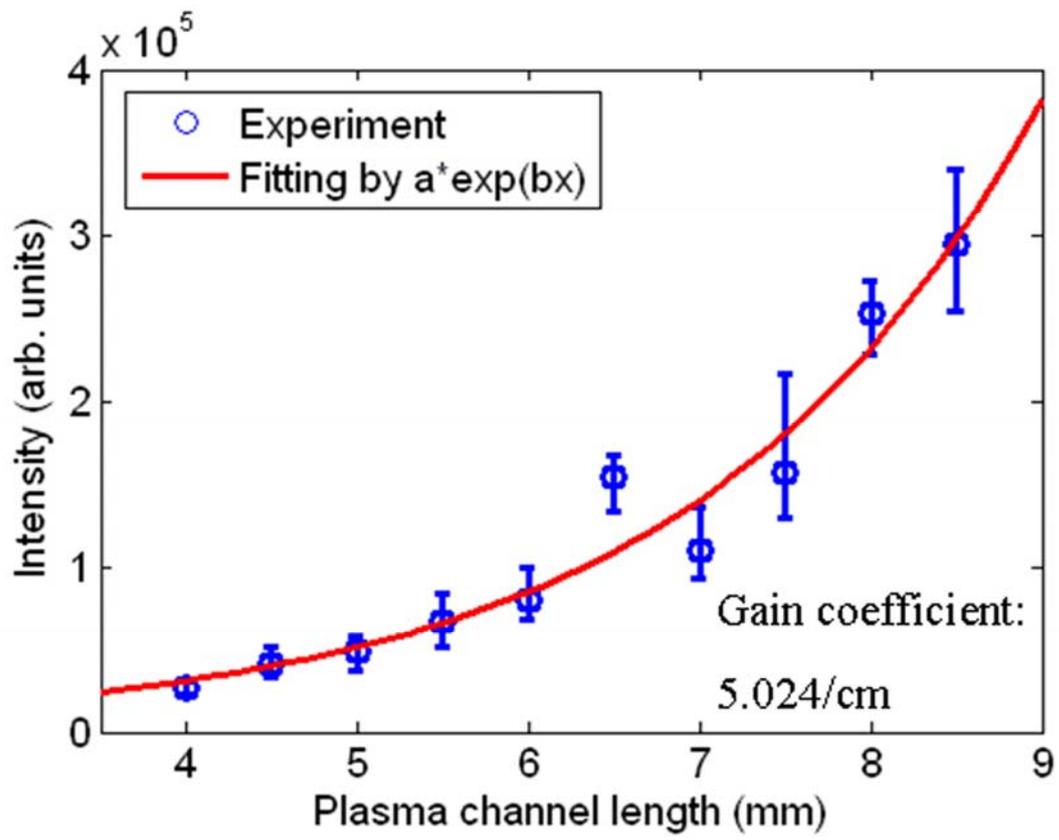

Fig. 5

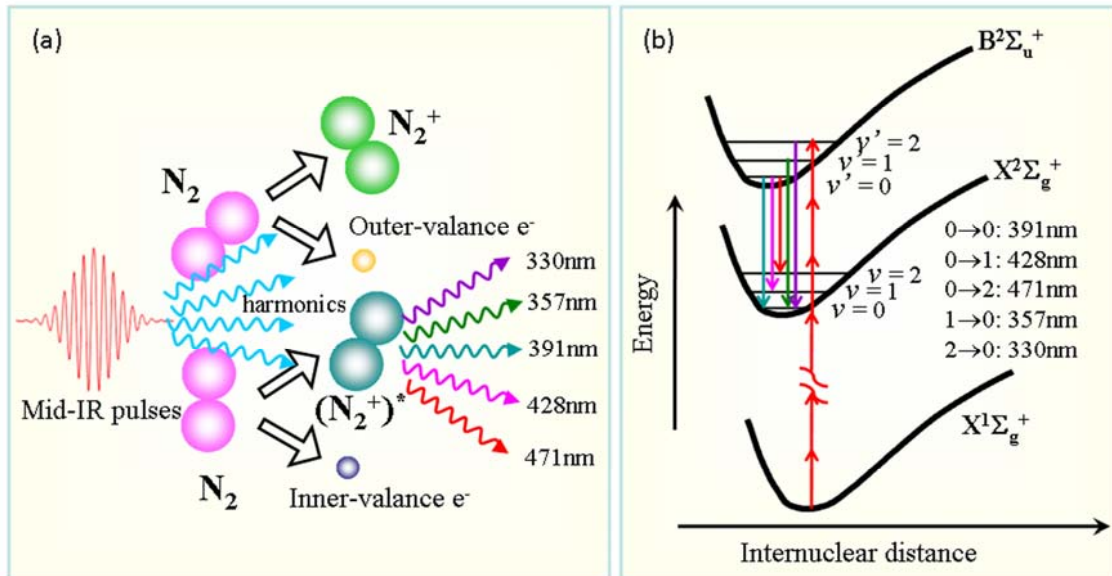